\documentclass[a4paper]{jpconf}
\usepackage{graphicx}
\usepackage{amsmath}
\usepackage{amssymb}
\usepackage{subcaption} 

\begin{document}

\title{Transport with hard-core interaction in a chain of asymmetric cavities.}

\author{G. P. Su\'arez$^{1,2}$, M. Hoyuelos$^{1,2}$, H. O. M\'artin$^{1,2}$}

\address{$^{1}$Instituto de Investigaciones F\'isicas de Mar del Plata (IFIMAR -- CONICET)}
\address{$^{2}$Departamento de F\'isica, Facultad de Ciencias Exactas y Naturales,
              Universidad Nacional de Mar del Plata. De\'an Funes 3350, 7600 Mar del Plata, Argentina}

\ead{gsuarez@mdp.edu.ar}

\begin{abstract}
  In this paper we investigate the diffusion of particles inside a chain of asymmetric cavities. We are considering particles that interact through a hard--core potential and are driven by an external force.
  We show that the difference in the current when the force is applied to the left and to the right strongly depends on the concentration inside the cavity.
  We found that, when the concentration is high enough, the hard--core interaction vanishes and inverts the asymmetric effect of the cavity. We also introduce a new equation, a modification to the Fick--Jacobs equation, to describe this system analytically.
  Finally, we used numerical simulations to verify the analytic results, finding a good agreement between theory and simulations.
\end{abstract}

\section{Introduction}
The diffusion of particles is a subject that has been studied extensively in many different contexts (see \cite{libro-diff-1} and cites within). One of this contexts is the diffusion of particles inside narrow channels \cite{rubi2010, hanggi2009, burada2008, Ai2006, dagdug2012}. However, the diffusion of interacting particles inside a confined environment has received much less attention, even when there are numerous examples in nature \cite{pries96, hedlishvili2001, zhou2008, karger, keil, beerdsen}.

In this paper we studied the transport of Brownian particles in a channel of variable transverse width. On the one hand, we studied how the current of particles that cross an asymmetric cavity is modified by increasing the concentration, considering particles that interact through a hard-core potential. On the other hand, we introduced a modified Fick--Jacobs equation to describe this process.
This equation has been solved using numerical methods for the case of asymmetric cavities. We mainly studied how the particles arrange inside the cavity, when a small force is applied. Numerical simulations with the method of Monte Carlo where carried out. The results obtained with the equation are in complete agreement with the numerical simulations.

This paper is organized as follows.
  In Sec.~\ref{sec:triangle}, we analyze the current of particles in a triangular cavity as a function of the concentration and we expose some of the characteristics of the numerical simulations.
  In Sec.~\ref{sec:equation}, we explain the most important steps to derive the non--linear Fick--Jacobs equation.
  In Sec.~\ref{sec:results}, we show how the theoretical solution matches with the results obtained with the numerical simulations.
  Some final remarks are stated in Sec.~\ref{sec:conclusion}.

\section{Current of particles}
\label{sec:triangle}

\begin{figure}
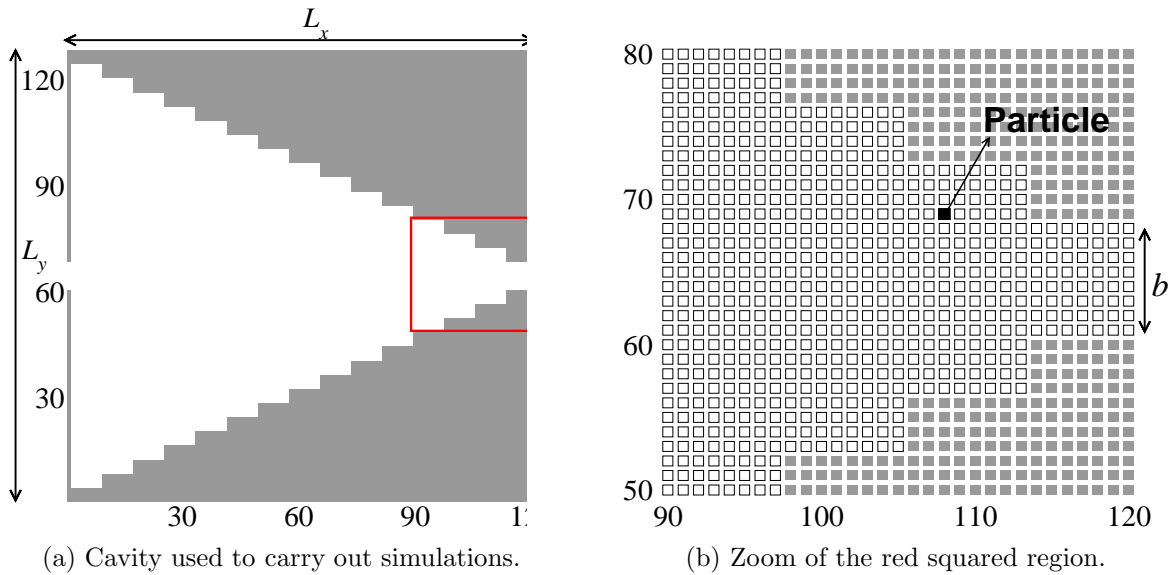

  \centering
  \begin{subfigure}[b]{0.45\textwidth}
	  \includegraphics[width=\textwidth]{{cavidad-triangular}.eps}
	  \caption{Cavity used to carry out simulations.}
	  \label{fig:triangulo}
  \end{subfigure}%
  \qquad 
  \begin{subfigure}[b]{0.45\textwidth}
	  \includegraphics[width=\textwidth]{{zoom-cavidad-triangular}.eps}
	  \caption{Zoom of the red squared region.}
	  \label{fig:zoom}
  \end{subfigure}
  \caption{Asymmetric cavity used in simulations.  The white region represents the allowed sites where the particles can diffuse. Distances are expressed in terms of the lattice constant, $a$. $L_x=120; L_y=128; b=8$.}
  \label{fig:CavidadTriangular}
\end{figure}

In the first part of this work we analyzed the diffusion of hard--core interacting particles inside a triangular cavity. 
To perform the numerical simulations we depicted the triangular cavity inside a regular square lattice of $L_x\times L_y$ sites, as seen in Fig.~\ref{fig:CavidadTriangular}. The distance between neighboring sites is $a$ (a value $a=1$ was considered). There is a small exit of size $b$ on each side. Periodic boundary conditions were set in those exits to simulate a long chain of identical cavities. Each site inside the cavity can be either occupied by a particle or empty. The hard--core interaction is realized by allowing only one particle on each site of the lattice. 

We define the mean concentration $ c = N / S$, where $S$ is the total number of sites inside the cavity and $N$ is the number of particles. At the beginning of the simulation, the particles are distributed randomly inside the cavity. The size of a particle is exactly one site, and they move by jumping to one nearest neighbor site. They are allowed to jump freely inside the cavity, except when the new site is occupied with another particle. In this case the particle remains on its site. We say that one Monte Carlo step occurs when, in average, every particle has had the opportunity to jump once. 

The particles are driven by a dimensionless external force, $\delta=\beta F a$ where \mbox{$\beta^{-1}=k_B T$}. At any given time, a particle has four possible directions to move. One of the them has a a higher jump rate, due to the effect of the external force applied. The other three have the same (lower) jump rate. Thus, if the force is applied to the right, the jump rates in different directions are
\begin{equation}
  \begin{split}
    p_\uparrow = p_\downarrow = p_\leftarrow = P \\
    p_\rightarrow = P(1+\delta),
  \end{split}
  \label{eq:rate1}
\end{equation}
where $P=D/a^2$, and $D$ is the diffusion coefficient. In the same way, if the force is applied to the left \mbox{$p_\leftarrow = P(1+\delta)$}, and $p_\rightarrow = P$. In the first part of this paper we will analyze the difference in the current of particles of a system when the force is applied first to the right, and then to the left. 
We let the system evolve to the stationary state, and then we start to measure the relevant quantities, like the distribution of the particles inside the cavity, the distance covered by each one or the mean velocity, among others.

Here we are interested in the total current of particles that cross the cavity, $j$. If there were no interaction between particles, the current will be higher as the concentration or the force becomes higher. But that is not necessarily true for interacting particles.
Also, because of the particular shape of the cavity, the current will not be the same when the external force is applied to the right, $j_+$, or to the left, $j_-$.
Since this asymmetric effect depends on the size of the particles, it has been proposed by Reguera \textit{et al.}\cite{Reguera2012} as a possible way to separate a mixture of different size particles. 
\begin{figure} 
  \centering
  \includegraphics[width=0.70\textwidth]{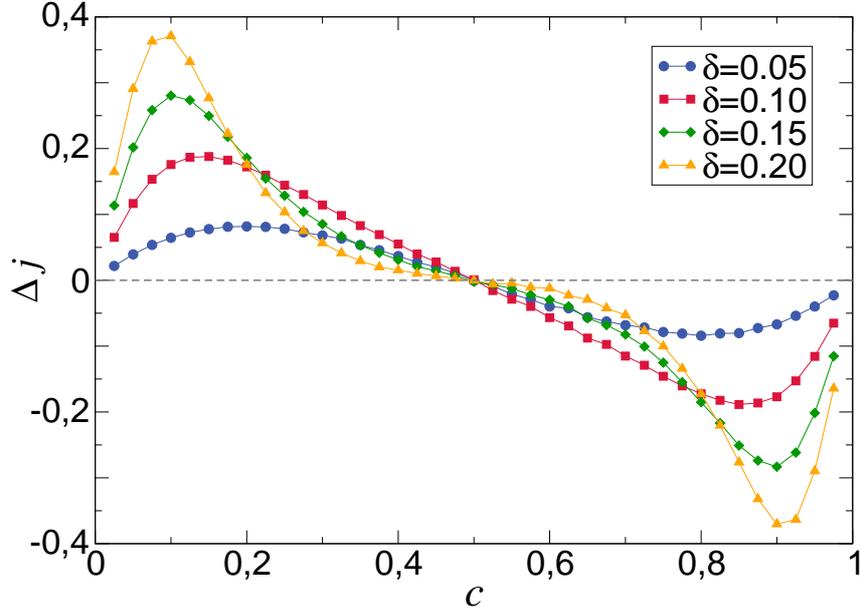}
  \caption{Difference in the current, $\Delta j$ a a function of the mean concentration $c$ in the cavity of Fig.~\ref{fig:CavidadTriangular}. The values of the force are: $\delta=0.05 (\bullet); \delta=0.10 (\blacksquare); \delta=0.15 (\blacklozenge); \delta=0.20 (\blacktriangle)$. Average over $10^6$ samples.}
  \label{fig:DeltaJ}
\end{figure}

In Fig.~\ref{fig:DeltaJ} we show the difference of currents of particles, $\Delta j = |j_+| - |j_-|$ as a function of the concentration for a wide range of forces. There are a few things that we can notice from this plot. Around $c=0.1$ we can see there is a maximum value in the difference of the current $\Delta j$. There is a value of the mean concentration, $c$, for each value of the force, $\delta$, for which the difference in the current is maximum. Increasing the concentration after this point, will only decrease the asymmetric effect of the cavity. Particularly, if the mean concentration is $c=0.5$ the asymmetric effect vanishes, and the current of particles is the same when the force is applied to the left or to the right, i.e. $\Delta j=0$, independently of its magnitude. Finally, if the concentration is higher than $0.5$, we found that the current of particles when the force is applied to the left is always higher than the current of particles when the force is applied to the right.
This is specially interesting because the asymmetry in the cavity is designed to favor the diffusion to the right. After this we can conclude that the hard--core interaction, whose influence increases as $c$ is increased, is the cause of an inversion of the asymmetry effect.

\section{Fick-Jacobs Equation}
\label{sec:equation}

The Fick-Jacobs equation describes the diffusion of non--interacting particles inside a channel of variable width. It has been used to describe the diffusion of not--interacting particles or in the low concentrations regime \cite{jacobs,zwanzig}. In this work we have extended this equation to include the case of diffusion of particles with hard--core interaction. With this we mean that two particles can not occupy the same space. The excluded volume effect of the other particles has a considerable influence on the diffusion of a tagged particle. 
This equation allows us to describe a rather complex system, of many interacting particles inside a two-- or three--dimensional environment, with a simple uni-dimensional equation.

To derive the modified Fick--Jacobs equation, let us consider a set of hard--core interacting particles in a one--dimensional discretized system, the so called single file diffusion \cite{harris, richards, beijeren}. In a given configuration, the current of particles between the sites $i$ and $i+1$ is given by,
\begin{equation}
  \begin{split}
    J_i^o &= n_i^o P_{i,i+1} - n_{i+1}^o P_{i+1,i}\\
	  &= P (1-\beta \Delta U_i)\, n_i^o\, (1 - n_{i+1}^o) - P \, n_{i+1}^o\, (1 - n_i^o),
    \label{curr2}
  \end{split} 
\end{equation}
where $n_i^o=0\: \mathrm{or}\: 1$ is the occupation number of site $i$, $\Delta U_i=U_{i+1}-U_i$ is the variation of the external potential and $\beta^{-1}=k_B T$.  $P$ is the jumping rate, $P=D/a^2$ where $D$ is the diffusion constant and $a=1$ is the length of the jump. We are assuming that $\beta |\Delta U_i| \ll 1$. The terms $(1-n_{i+1}^o)$ and $(1-n_i^o)$ take into account the hard--core interaction.

Averaging over configurations one has $n_i=\langle n_i^o\rangle$ and $J_i=\langle J_i^o\rangle$. In the continuous limit $n_i \rightarrow n(x_i)$, $U_i \rightarrow U(x_i)$ and $J_i \rightarrow J(x_i)$, where $x_i=i$. Assuming the decorrelation $\langle n_i^o n_{i+1}^o \rangle \simeq n(x_i) n(x_{i+1})$ and small variation of $n(x)$ we obtain

\begin{equation}
 J=-D \left[ \frac{\partial n}{\partial x}+\beta \frac{dU}{dx} n (1-n) \right] .
 \label{eq:J}
\end{equation}

\begin{figure}
	\centering
	\begin{subfigure}[c]{0.45\textwidth}
		\includegraphics[width=\textwidth]{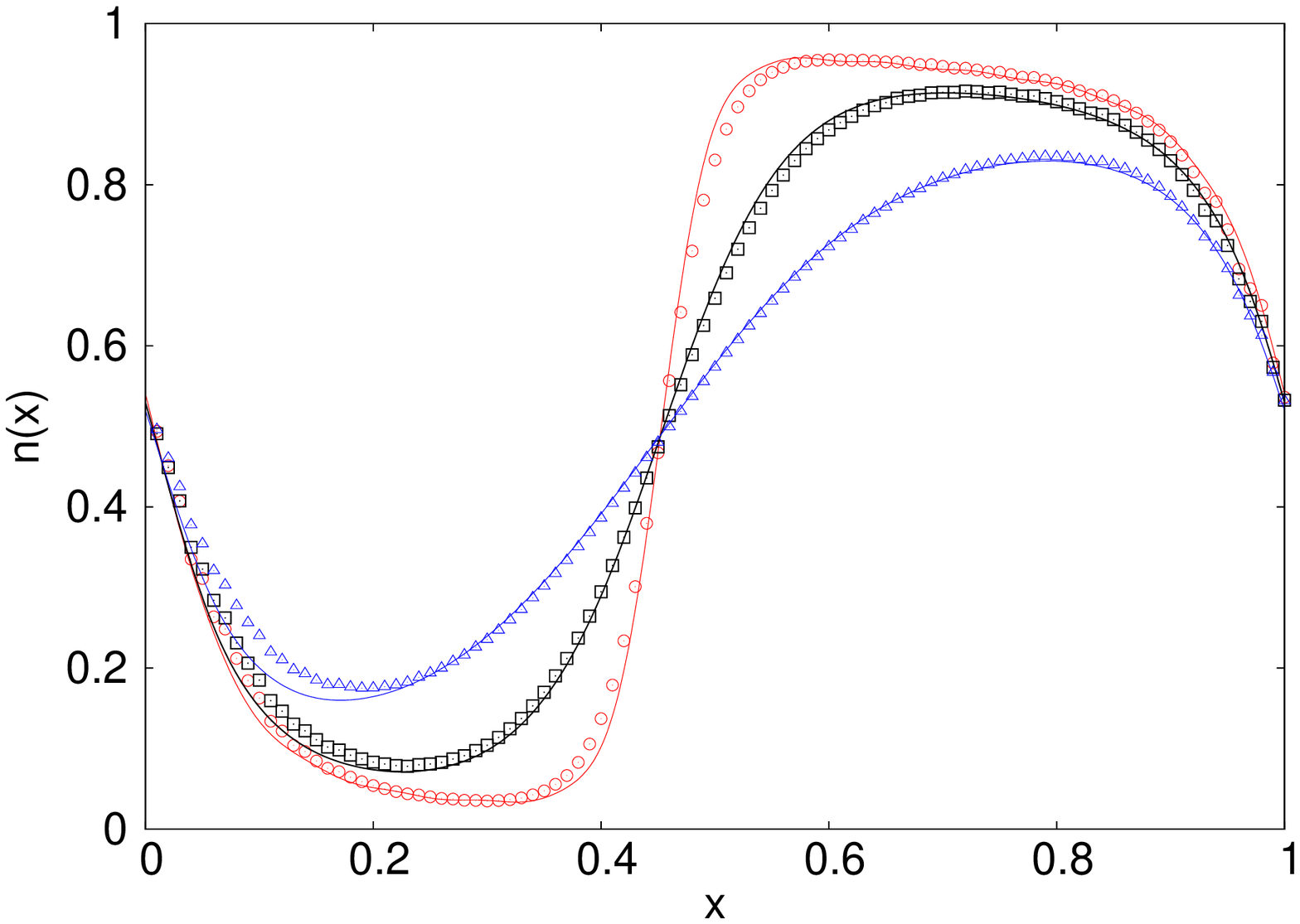}
		\caption{ }
		\label{fig:fj}
	\end{subfigure}%
	~ 
	\begin{subfigure}[c]{0.53\textwidth}
		\includegraphics[width=\textwidth]{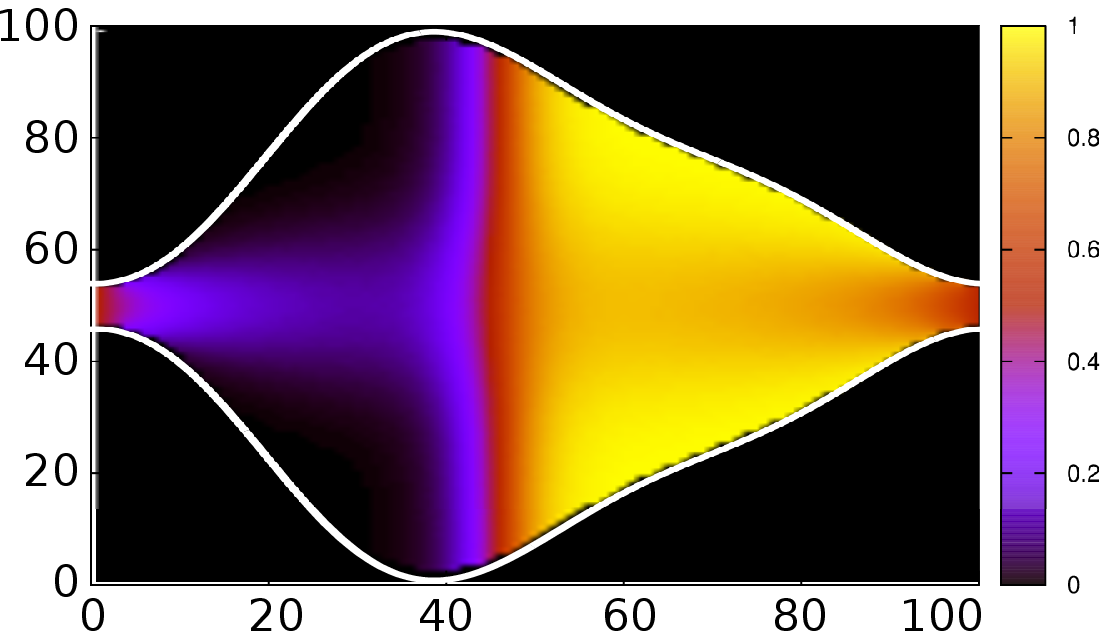}
		\caption{ }
		\label{fig:color}
	\end{subfigure}
	\caption{
	(a) Concentration along the $x$--axis, $n(x)$. Comparison between analytic solution to Eq.~(\ref{eq:final}) and simulations for $c=0.5$. In each case, the force is applied to the right and the following values were used: $\delta=0.1$ (blue triangles), $\delta=0.2$ (black squares) and $\delta=0.5$ (red circles). Dots were obtained with numerical simulations ($10^5$ samples between $t=10^4$ and $t=10^7$ Monte Carlo time-steps).
	(b) Site concentration inside the cavity. Corresponds to the case $c = 0.5$ and $\delta=0.5$ (red circles in (a)).
	}
	\label{fig:comparison}
\end{figure}

Let us now consider a two--dimensional channel composed by a chain of cavities (as the one shown in Fig.~\ref{fig:color}) where the external potential $U(x)$ is independent of the transverse direction, $y$. Assuming local equilibrium, the site concentration $m(x,y)$ does not depend on the transverse direction either. That is, $m(x,y)=n(x)$. Then, the total current is $j=JA$, where $A(x)$ is the transverse width of the channel and from Eq.~(\ref{eq:J}) one obtains
\begin{equation}
  -\frac{j}{D} = A \left[ \frac{\partial n}{\partial x} - \beta F n (1-n) \right] ,
   \label{eq:final}
\end{equation}
where $F=-dU/dx$ is the external force. Now from the continuity equation $\frac{\partial m}{\partial t} + \frac{\partial j}{\partial x} = 0$ we get

\begin{equation}
 \frac{\partial m}{\partial t} = D \frac{\partial}{\partial x}\left[ \frac{\partial m}{\partial x} - \frac{d A}{d x} \frac{m}{A} - \beta F m \left(1 - \frac{m}{A}\right) \right] , 
\end{equation}
which corresponds to our extension of the Fick-Jacobs equation where the hard-core interaction is taken into account. For non--interacting particles $(1-\frac{m}{A}) \rightarrow 1$ and the well known Fick--Jacobs equation is recovered. 

\section{Results and Discussion}
\label{sec:results}

To test how accurate Eq.~(\ref{eq:final}) is, we solved it for a particular case using a smoother asymmetric cavity than the one shown in Fig.~\ref{fig:CavidadTriangular}, see Fig.~\ref{fig:color}. 

In Fig.~\ref{fig:fj} we show the stationary state of site concentration along the direction of the channel in an asymmetric cavity, for different values of the force. In the three curves the force is applied to the right. 
However, as a consequence of the analogy in the definition of a particle and a vacancy, they also represent the system behavior when the force is applied to the left. For every particle that move to the right there is a vacancy that moves to the left. All we have to do is to change  $n \to 1-n$ (then $c \to 1-c$), $\delta \to -\delta$ and $j \to -j$. Note that in Eq.~(\ref{eq:final}) this symmetry holds, see also Fig.~\ref{fig:DeltaJ}. Further details on this matter may be found in \cite{suarez-cavidad,dierl2013,torrez2013}.

The results show that the agreement between theory and simulations is very good almost everywhere. However, it can be noticed that, in some regions the agreement is better than others. This is due to the fact that we assumed $m$ to be constant along the $y$ axis. When this condition is not fulfilled, the theoretical curves slightly differs from simulations (see Fig.~\ref{fig:comparison}). 

\section{Conclusion}
\label{sec:conclusion}
In this paper we studied how the diffusion of interacting particles is modified by increasing the concentration. We showed that the interaction between particles plays a fundamental role in this kind of systems. One reason is the formation of clogs where the available space is limited.

Apart from that, we were able to obtain a non--linear Fick-Jacobs equation to describe the transport of hard--core particles inside a cavity with variable cross section. This equation correctly predicts the steady state of the concentration of particles along the direction of the channel. 

The agreement of theory and simulations confirms that the modified equation of Fick--Jacobs here introduced is a useful tool to describe the transport of particles when the interactions between them can not be neglected. 

\section*{Acknowledgments}
This work was partially supported by Consejo Nacional de Investigaciones Cient\'{\i}ficas y T\'{e}cnicas (CONICET, Argentina, PIP 0041 2010-2012). GPS wishes to thank the organizing committee of the XXVI IUPAP Conference on Computational Physics for financial support to assist to the conference.

\end{document}